\documentclass[
aps,%
12pt,%
final,%
notitlepage,%
oneside,%
onecolumn,%
nobibnotes,%
nofootinbib,%
superscriptaddress,%
noshowpacs,%
centertags]%
{revtex4}

\usepackage{epsfig}
\usepackage{enumerate}
\usepackage{longtable}

\begin{document}

\pagenumbering{arabic}

\title{New set of experimental wavenumber values for visible part of emission spectrum 
($545 \div 627$ nm) of the $D_2$ molecule with partly resolved fine structure of 
triplet-triplet rovibronic transitions}

\author{B.~P.~Lavrov}
\email{lavrov@pobox.spbu.ru}
\author{I.~S.~Umrikhin}
\affiliation{
Faculty of Physics, St.-Petersburg State University, \\
St.-Petersburg, 198504, Russia}

\begin{abstract}

New set of wavenumber values for electronic--vibro--rotational (rovibronic) transitions in limited 
part of the emission spectrum of the $D_2$ molecule ($545 \div 627$ nm) have been measured with a precision 
$0.007 \div 0.1$ cm$^{-1}$ depending on the translational temperature in plasma, the signal--to--noise ratio, 
and the degree of overlap with adjacent spectral lines. 
For the first time partly resolved fine structure of the $D_2$ spectral lines 
has been observed in the visible part of the spectrum. 

\end{abstract}

\maketitle

\section*{Introduction}

Experimental studies of the $D_2$ spectrum have been started just after discovery of 
atomic deuterium \cite{Urey1932_1, Urey1932_2}. First reports were motivated by the 
problem of spectroscopic determination of the nuclear spin of deuterium 
\cite{LewisAshley1933, MurphyJohnston1934} and by appearance of the opportunity to get 
more information about structure of diatomic molecules $NH$ and $H_2$ from spectra of 
their isotopic species $ND$ \cite{DiekeBlue1933}, and $HD, D_2$ 
\cite{DiekeBlue1934, DiekeBlue1935, Dieke1935}. 
Later on studies of spectra and structure of molecular deuterium were stimulated 
not only by understandable general interest (an isotopomer of simplest 
neutral diatomic molecule --- natural touchstone for theoretical models), but because of 
direct practical value in connection with wide use of $D_2$ in physical experiments and in 
various applications. However, our knowledge of optical spectrum of molecular deuterium is 
still insufficient in spite of tremendous efforts by spectroscopists over the previous century. 
Up to now most of spectral lines have not yet been assigned. As an example, in the latest 
compilation of experimental data \cite{FSC1985} the working list of 27488 recorded lines 
(within the wavelength range $\approx309-2780$ nm) contains only 8243 assignments.

The band spectrum of the $D_2$ molecule is caused by both singlet--singlet 
and triplet--triplet radiative electronic--vibro--rotational (rovibronic) 
transitions\footnote{Well known and rather important feature of emission spectra 
of diatomic hydrogen isotopomers --- wide ($160-500$ nm) continuum due to spontaneous 
transitions from vibro--rotational levels of the bond $a^3\Sigma_g^+$ electronic state 
to the repulsive $b^3\Sigma_u^+$  state --- is out of the scope of present paper because 
it can't be used for determination of rovibronic term values.}. The intercombination 
lines have not yet been observed. 

The most interesting resonance singlet band systems are located in vacuum ultraviolet (VUV). 
Measurements of wavenumbers of separate rovibronic lines and empirical determination of singlet 
rovibronic term values is in progress up to now 
\cite{RLTBjcp2006, RLTBjcp2007, RIVLTUmol2008, LSJUMchp2010, GJRT2011, DIUROJNTGSKEmol2011}. 
The precision of wavenumber measurements for the $D_2$ rovibronic lines in VUV is now close to 
$0.05-0.1$ cm$^{-1}$ \cite{DIUROJNTGSKEmol2011} by conventional methods, while a laser technique 
achieved unprecedented accuracy of $\approx 0.006$ cm$^{-1}$ \cite{RIVLTUmol2008}.

The triplet transitions are responsible for a major part of light emission of 
ionized gases and plasma in near infrared, visible and near 
ultraviolet\footnote{Bands located in visible part of spectrum are especially 
interesting because they are often used for spectroscopic diagnostics of non-equilibrium 
plasmas (see e.g. \cite{RDHL2008}).}. All empirical data concerning wavenumbers and rovibronic 
term values of $D_2$ obtained by means of emission, absorption, Raman and anticrossing 
spectroscopy were collected, analyzed and reported in \cite{FSC1985}. Since that time very 
few new experimental data about triplet rovibronic transitions of $D_2$ were obtained by 
Fourier transform infrared (FTIR) \cite{DabrHerz}, IR tunable laser \cite{Davies}, 
and emission \cite{LU2008, LU2009} spectroscopy. It should be noted that at present almost 
all available experimental data on rovibronic line wavenumbers of the $D_2$ molecule were 
obtained by photographic recording of emission or absorption spectra.

Our recent studies \cite{LU2008, LU2009} revealed that wavenumber values for triplet rovibronic 
transitions reported in \cite{FSC1985} have significant differences from values predicted by 
Rydberg-Ritz combination principle and our own data obtained by photoelectric recording 
\cite{LU2009}\footnote{The same situation was earlier observed for rovibronic 
transition wavelength values from \cite{Dieke1972} in the $H_2$ spectrum, 
see e.g. the spread of experimental points on the fig.3 in \cite{ALMU2008}.}.
The minority of the differences is caused by misprints and erroneous 
line assignments. But vast majority of the differences are about 
$0.01 \div 0.1$ cm$^{-1}$ and show random spread 
around "synthesized" wavenumber values, calculated as differences of optimal 
energy level values from \cite{LU2008}. We suppose that they appear due to 
a finite precision in reading from photo plates by 
microphotometric comparators, round-up errors 
in calculating the wavenumber values from measured in air wavelengths, 
and shifts of the photographic density maxima for blended lines relative to 
an actual position of the line centers. 
This random spread of available wavenumber values together with absence of reliable error 
bars for each value seriously limited determination of rovibronic term values by means of 
optimization method \cite{LR2005} when it was applied for an analysis of triplet rovibronic 
transitions of the $D_2$ molecule in \cite{LU2008}.

Therefore we decided to start systematic studies of visible part of emission spectra of the $D_2$ 
molecule for obtaining more precise and more reliable wavenumber values of rovibronic transitions. 
The present paper reports first results for certain limited part of the spectrum.

\section*{Experimental}

We used experimental setup described elsewhere \cite{ALMU2008, LMU2011}. 
Emission of plasma inside molybdenum capillary located between anode and cathode of 
a gas discharge tube was used as a light source. The flux of radiation through a 
hole in an anode was focused by achromatic lens on the entrance slit of the spectrometer. 
Detailed description of the self--made high resolution automatic spectrometer and 
corresponding software was reported in \cite{LMU2011}. The $2.65$ m Ebert-Fastie spectrograph 
with $1800$ grooves per mm diffraction grating was equipped with additional camera lens 
(that gives effective focus length $F=6786 \pm 8$ mm) and computer-controlled CMOS matrix 
detector ($22.2 \times 14.8$ mm$^2$, $1728 \times 1152$ pixels). The apparatus has linear dispersion 
of $0.076 \div 0.065$ nm/mm in the wavelength range $400-700$ nm, dynamic range of measurable 
intensities greater than $10^4$ and maximum resolving power up to $2 \times 10^5$. However, actual 
resolving power in our conditions was mainly limited by Doppler broadening of the $D_2$ spectral 
lines due to small reduced mass of nuclei.

It should be emphasized once more that the overwhelming majority of data on the wavenumbers 
for rovibronic transitions of the $D_2$ molecule was obtained by photographic recording of spectra. 
Our way of determination of the rovibronic transition wavenumbers developed in \cite{ALMU2008, LU2009, LMU2011} 
is based on linear response of the CMOS matrix detector on the spectral irradiance and digital intensity recording. 
Both things provide an extremely important
advantage of our technique over traditional photographic recording 
with microphotometric comparator reading. It
not only makes it easier to measure the relative spectral line
intensities but also makes it possible to investigate the shape
of the individual line profiles and, in the case of overlap of
the contours of adjacent lines (so--called blending), to carry
out numerically the deconvolution operation (inverse to the convolution
operation) and thus to measure the intensity and wavelength
of even blended lines. As is well known, it is this blending
that makes it very hard to analyze dense multiline spectra of 
molecular hydrogen isotopomers \cite{Dieke1972, FSC1985}

It is known that, in the case of long--focus spectrometers, the dependence of the wavelength 
on the coordinate $x$ along direction of dispersion, is close to linear in the vicinity of the center
of the focal plane. It can be represented as a power series expansion over of the small 
parameter $x/F$, which in our case does not exceed $2 \times 10^{-3}$.
\footnote{The $x$--coordinate actually represents small displacement from the center 
of the matrix detector. $F$ is the focal length of the spectrometer mirror.}
On the other hand, the 
wavelength dependence of the refractive index of air $n(\lambda)$ is also close to linear 
inside a small enough part of the spectrum. Thus, when recording narrow spectral intervals, 
the product $\lambda_{vac}(x) = \lambda(x) n(\lambda(x))$ has the form of a power series of low degree. 
This circumstance makes it possible to calibrate the spectrometer directly in vacuum wavelengths 
$\lambda_{vac} = 1 / \nu$, thereby avoiding the technically troublesome problem of accurate 
measuring the refractive index of air under the various conditions under which measurements 
are made. 

Another peculiarity of our calibration technique is using of experimental vacuum wavelength values 
from \cite{FSC1985} as standard reference data. We already mentioned above that those data show small 
random spread around smooth curve representing dependence of the wavelengths on positions of corresponding 
lines in the focal plane of the spectrometer (see e.g. \cite{ALMU2008}. Moreover those random errors are 
in good accordance with normal Gaussian distribution function. Thus it is possible to obtain precision 
for new wavenumber values better than that of the reference data due to smoothing. 

To be sure that the data from \cite{FSC1985} are free from systematic errors we have had to perform 
special experiments with capillary--arc lamp analogous to that described in \cite{LSh1979} 
(capillary diameter $d = 1.5$ mm and current density $j = 30$ A/cm$^2$) but filled with the $H_2+D_2+Ne$ 
mixture (1:1:2) under total pressure $P \approx 8$ Torr.

For vacuum wavelength calibration we used bright free of blending lines of the $D_2$ 
and $H_2$ molecules as well as $Ne$ spectral lines with reference data from \cite{FSC1985, Dieke1972, SS2004} 
respectively. 

\begin{figure}[!ht]
\begin{center}
\epsfig{file=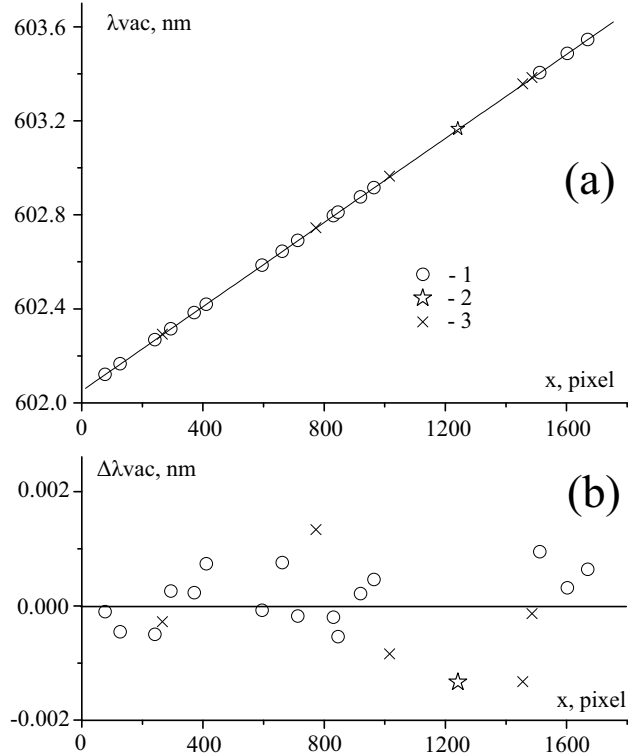, width=0.5\columnwidth,clip}
\end{center}
\caption{Fragment of dependences of vacuum wavelengths $\lambda_{vac}$ (a) and the differences 
$\Delta \lambda_{vac}$ (b) of the brightest $D_2$, $H_2$ and $Ne$ spectral lines on the coordinate 
(in pixels) in the focal plane of the spectrometer. 
Points $1$ --- the values for $D_2$ molecule from \cite{FSC1985};
Points $2$ --- the values for $Ne$ atom from \cite{SS2004};
Points $3$ --- the values for $H_2$ molecule from \cite{Dieke1972}.
Solid line represents the approximation of experimental data.}\label{d2h2ne}
\end{figure}

As an example the dependence of vacuum line wavelength
on its position on CMOS matrix in pixels for strong unblended lines is shown on fig.\ref{d2h2ne}(a) for
wavelength region corresponding to size of CMOS matrix.
One may see that the dependence of the wavelengths for most of the lines on the coordinate is
monotonic and close to rectilinear. 
The calibration curve of the spectrometer was obtained by the polynomial least--squares fitting of the data.
Our measurements showed that, using a linear hypothesis is inadequate and a third--degree polynomial
is excessive, while an approximation by a second--degree polynomial provides calibration accuracy 
better than $2 \times 10^{-3}$ nm.
Such a wavelength calibration allows us to get new experimental values for the rovibronic line wavenumbers.
The differences $\Delta \lambda_{vac}$ between the new values and used reference data are shown 
in fig.\ref{d2h2ne}(b).
One may see that the differences have certain spread around calibration curve, that does not exceed $0.002$ nm.
Thus our measurements show that experimental wavenumber values from \cite{FSC1985, Dieke1972, SS2004} 
are in good agreement with each other. 
Therefore in our studies of the $D_2$ spectrum the vacuum wavelength values from \cite{FSC1985} were used as
the reference data set.

Such "internal reference light source" gave us an opportunity to eliminate experimental wavenumber errors 
caused by the shift between a spectrum under the study and the reference spectrum from another reference 
light source, due to a different illumination of the grating by the different lamps (see e.g.\cite{RLTBjcp2006}). 

\section*{Results and discussion}

For small regions of the spectrum ($\approx 0.5$ nm wide)\footnote{That corresponds 
to one third of the matrix: $550-600$ pixels wide.}
the observed spectral intensity distribution was approximated by superposition of a 
certain number of Gauss or Voigt profiles with the linewidth $\Delta \nu_{obs}$\footnote{
We use usual meaning for a linewidth, namely full width on half maximum (FWHM). In our case
observed linewidth $\Delta \nu_{obs}$ includes both instrumental profile and that of 
broadening in plasma, mainly due to Doppler effect.} equal for all the 
profiles within a spectral region under the consideration. Thus, optimal values for adjustable 
parameters (line centers, relative intensities and one common value of $\Delta \nu_{obs}$ for the region) 
were obtained by solving the reverse spectroscopic problem in the framework of maximum--likelihood 
principle by means of special computer program based on Levenberg--Marquardt's 
algorithm \cite{L1944, M1963}. 

Determination of wavenumber values for line centers (wavenumbers of rovibronic transitions) by means 
of the deconvolution process described above gives us an opportunity to reach much higher 
resolution than that predicted by the Rayleigh criterion. This fact may be illustrated by the 
example shown in fig.\ref{3exp}. It represents experimental intensity distributions (hollow circles) for the 
same narrow wavenumber range measured under three different conditions:

\begin{figure}[!ht]
\begin{center}
\epsfig{file=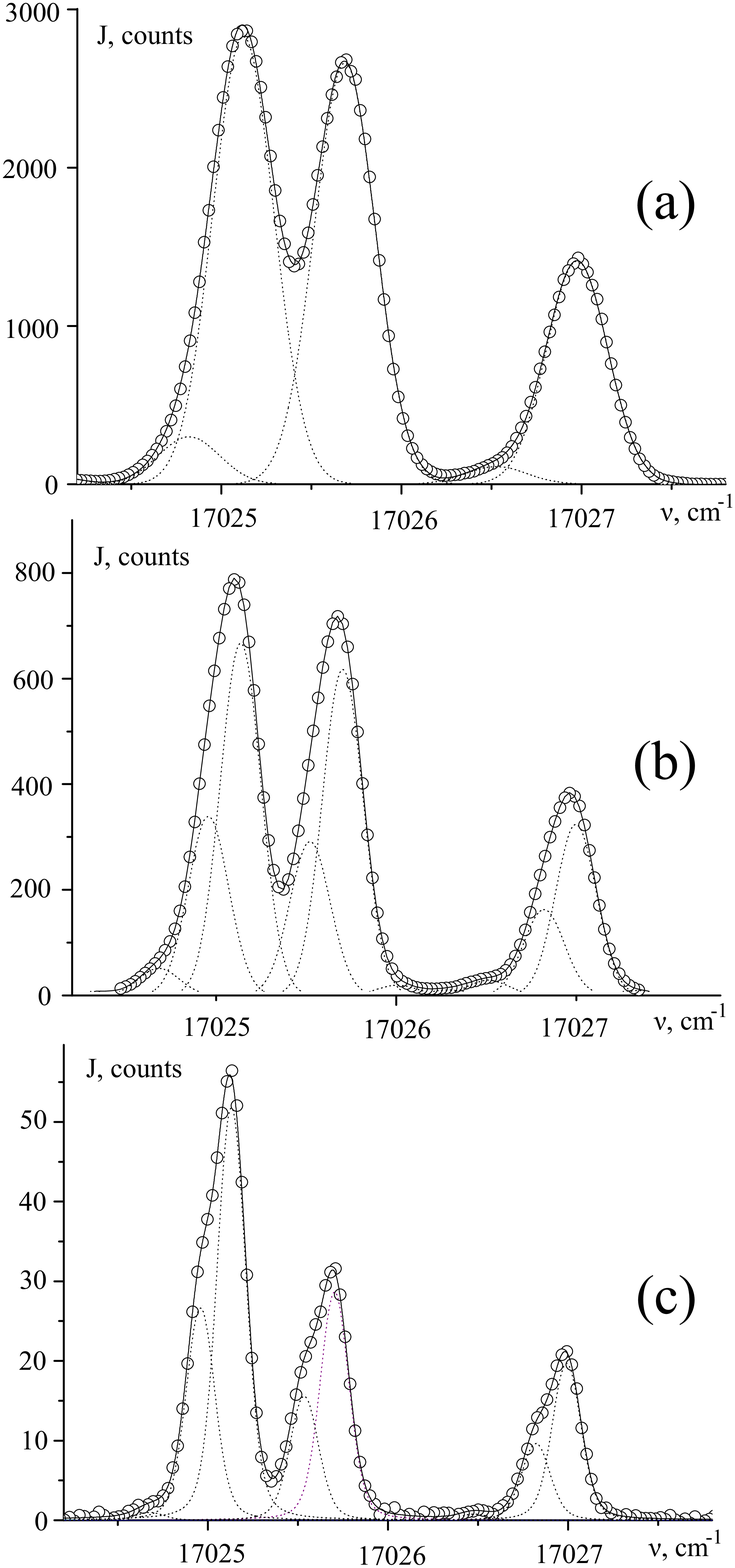, width=0.45\columnwidth,clip}
\end{center}
\caption{Fragment of the $D_2$ spectrum in the spectral range 
$17024-17028$ cm$^{-1}$ containing $R4$, $R5$ and $R6$ spectral lines for the $(1-1)$ 
band of the $i^3\Pi_g \to c^3\Pi_u$ electronic transition. 
Experimental intensity $J$ in counts is shown by open circles. 
Dotted lines represent Gaussian (a), (b) and Voigt (c) profiles for separate lines obtained by deconvolution, 
while the solid line corresponds to the total intensity obtained by summing over the components.
Cases (a), (b), and (c) correspond to experimental conditions (a), (b), and (c) (see text).}\label{3exp}
\end{figure}

\begin{enumerate}[(a)]
\item Hot cathode capillary--arc discharge lamp LD2-D \cite{GLT1982} under current density $j = 10$ A/cm$^2$ and 
gas temperature in plasma $T = 1500 \pm 150$ K\footnote{The temperature was obtained from the intensity distribution in
Q-branch of (2-2) band of $d^3\Pi_u^- \to a^3\Sigma_g^+$ electronic transition.} 
(large Doppler width, $\Delta \nu_D  = 0.22$ cm$^{-1}$) and the 
entrance slit width $\Delta X = 60$ $\mu m$ four times over than the so-called normal width 
(large instrumental profile);

\item the same conditions in plasma as those for case (a), but the slit width $\Delta X = 15$ $\mu m$ close 
to normal (providing optimal width of the instrumental profile);

\item cold cathode glow discharge in water cooled quartz tube under $j = 0.4$ A/cm$^2$, $T = 640 \pm 50$ K\footnote{
The temperature was obtained both from the intensity distribution in Q-branch of (2-2) band of 
$d^3\Pi_u^- \to a^3\Sigma_g^+$ electronic transition and from Doppler broadening of spectral lines.}
(smaller Doppler width $\Delta \nu_D = 0.15$ cm$^{-1}$), and $\Delta X = 15$ $\mu m$.
\end{enumerate}

The observed line profiles for strong unblended lines in the cases (a) and (b) 
were close to the Gaussian shape except for insignificant far wings. Therefore, 
the intensity distribution was approximated by superposition of a certain number of Gaussian profiles.
In the case (c) the gas temperature is lower thus the observed line profiles 
were determined by Doppler and instrumental broadening. Therefore the intensity distributions 
obtained in the experiment (c) were fitted by the superposition of a certain number of
Voigt profiles.

As the result of the fitting we obtained following values of observed linewidths 
$\Delta \nu_{obs} = 0.38$, $0.27$, and $0.18$ cm$^{-1}$ for cases (a), (b), and (c) respectively.

One may see that when resolution of optical part of the spectrometer is insufficient (case (a))
then only 3 bright lines are distinguished. Decrease of $\Delta \nu_{obs}$ in the case (b) (due to more narrow
instrumental profile) makes it possible to observe that each of those lines consist of two
distinguishable components having the same intensity ratio close to $2$. Further decrease of
$\Delta \nu_{obs}$ in the case (c) (due to smaller Doppler broadening) leads to the same values
of the component wavenumbers and ratio of their intensities (see table \ref{tabdouplets}).

Joint analysis of splitting in such visible "doublets" (about $0.20$ cm$^{-1}$) and relative intensities of 
two main components of visible "doublets" (about $2.0$) show that they represent partly resolved 
fine structure of lines determined by triplet splitting of lower rovibronic levels of various 
$(1s\sigma nl\pi)^3\Lambda_g \to (1s\sigma 2p\pi)^3\Pi_u$ electronic transitions. 

Present paper reports the results of two main experiments corresponding to cases (a) and (c).
The results for wavelength range $545 \div 627$ nm are presented in Table \ref{tabnewlines}\footnote{
Spectral range for experiment (a) was $400 \div 700$ nm thus only a part of wavenumbers obtained 
in this experiment is presented in Table \ref{tabnewlines}. Therefore we used original 
numbering of spectral lines for that experiment, therefore the column $K_1$ does not begin with a unit.}.
One may see from the table that the wavelength values obtained in our experiments differ from those
collected in \cite{FSC1985} not only quantitatively but qualitatively as well.
We observed much more lines and part of them could be visible components of the fine structure of rovibronic lines.
Detailed analysis of the data is now in progress and will be reported elsewhere.

Separation of observed doublets ($0.17 \pm 0.01$ cm$^{-1}$) corresponds to 
data obtained by means of FTIR \cite{DabrHerz} and laser \cite{Davies} 
spectroscopy in infrared part of the spectrum.
The observed ratio of intensities of these doublets is close to that calculated by 
Burger--Dorgello--Ornstein sum rule (2.0).

Partly resolved fine structure of spectral lines provides the
opportunity to expand the existing identification of triplet
rovibronic lines by detecting those doublets in experimental spectra. 
Within the spectral region under the study ($545 \div 627$ nm) there are more than 
200 pairs of unassigned lines which may represent visible doublets of partly resolved 
triplet structure of rovibronic transitions between $^3\Lambda_g$  and  $c^3\Pi_u^-$ 
electronic states of the $D_2$ molecule. 

Obtained results reveal new opportunities for identifying the great number of 
currently unassigned the $D_2$ lines in visible part of the spectrum.

Present work was financially supported in part by the Russian
Foundation for Basic Research, Grant No. 10-03-00571-a.

\newpage

\begin{longtable}[]{cccccccc}
\caption[]{Wavenumber values $\nu$ (in cm$^{-1}$) of R4, R5 and R6 lines for the $(1-1)$ band of
$i^3\Pi_g^- \to c^3\Pi_u^-$ electronic transition for three experiments (a), (b) and (c) (see text). 
$N''$ --- quantum number of the total angular momentum of molecule 
excluding spins of electrons and nuclei for lower rovibronic level of the transition.
$\nu_s$ and $\nu_w$ --- wavenumbers for strong and weak components of visible doublet.
The uncertainties of the wavenumber determination (one SD) are shown in brackets in units 
of least significant digit.
}\label{tabdouplets}
\endfirsthead
\multicolumn{8}{l}{Table \ref{tabdouplets} (Continued).}\\ 

\hline
$N''$ & $\nu^{(a)}$ & $\nu^{(b)}_s$ & $\nu^{(b)}_w$ & $\nu^{(c)}_s$ & $\nu^{(c)}_w$ \\
\hline
\endhead

\hline
$N''$ & $\nu^{(a)}$ & $\nu^{(b)}_s$ & $\nu^{(b)}_w$ & $\nu^{(c)}_s$ & $\nu^{(c)}_w$ \\
\hline

4  & 17025.12(2) &  17025.14(3) & 17024.96(3) &	17025.13(3) &	17024.96(3) \\ 
5  & 17026.98(2) &  17027.00(3) & 17026.82(3) &	17026.99(3) &	17026.82(4) \\ 
6  & 17025.69(2) &  17025.70(3) & 17025.52(3) &	17025.70(3) &	17025.53(4) \\ 

\hline
\end{longtable}

\newpage

{\def\baselinestretch{1.05}
\setlength{\tabcolsep}{10pt}

\begin{longtable}[]{rlllrr}
\caption[]{Experimental wavenumber values (in cm$^{-1}$) for rovibronic spectral lines of 
$D_2$ molecule in $545 \div 627$ nm spectral region. 
$\nu^{(a)}$ --- hot cathode capillary-arc discharge experiment (a), 
$\nu$ --- data from \cite{FSC1985}, 
$\nu^{(c)}$ --- glow discharge experiment (c). 
The uncertainties of the wavenumber determination (one SD) are shown in brackets in units 
of least significant digit.
$K_1$ and $K_2$ --- line index for experiments (a) and (c).
Assignments are taken from the Appendix C of \cite{FSC1985}.}\label{tabnewlines}
\endfirsthead
\multicolumn{6}{l}{Table \ref{tabnewlines} (Continued).}\\ 

\hline
$K_1$ & $\nu^{(a)}$ & $\nu$ & $\nu^{(c)}$ & $K_2$ & Assignment \\
\hline
\endhead

\hline
$K_1$ & $\nu^{(a)}$ & $\nu$ & $\nu^{(c)}$ & $K_2$ & Assignment \\
\hline

&
& 18331.19
& 18331.28(5)	& 1
&  
\\
6296	& 18330.24(6)
& 18330.27
& 18330.316(15)	& 2
& S+  GK-2B  (3-5) P1 
 
\\
&
& 
& 18329.96(3)	& 3
&
\\
&
& 
& 18329.69(5)	& 4
&
\\
6297	& 18329.57(7)
& 18329.54
& 18329.53(3)	& 5
&  
\\
6298	& 18328.94(7)
& 18329.07
& 18329.082(18)	& 6
&  
\\
&
& 
& 18328.79(4)	& 7
&
\\
\hline
\end{longtable}
}

\end{document}